# NANOIONICS AND INTERFACES FOR ENERGY AND INFORMATION TECHNOLOGIES

Francesco Chiabrera[1], Iñigo Garbayo[1], Albert Tarancón[1,2]

[1] Department of Advanced Materials for Energy Applications, Catalonia Institute for Energy Research (IREC), Jardins de les Dones de Negre 1, 08930 Sant Adrià del Besòs, Barcelona, Spain.

[2] ICREA, Passeig Lluís Companys 23, 08010, Barcelona, Spain

## 1. Introduction to Nanoionics: Beyond Bulk Properties

There is a growing interest in the development of functional metal oxides with mixed ionic electronic conduction for their application in different strategic fields. In particular, ionic transport-related phenomena are of primary importance in energy transformation and storage devices, such as solid oxides fuel cells (SOFC) or batteries. Traditionally, the main issue that hindered a wide spread of solid state electrochemical devices is the slow ionic conduction and the reduced number of families of materials with pure ionic conductivity able to play the role of the electrolyte. This is due to the low mobility of the ionic charge carriers and their mechanical and coulombic interactions with their host crystal structures. For relatively big and charged ions, e.g. $O^{2-}$ or $Na^+$, the low ionic conductivity imposes operating at high temperatures, which is incompatible with some applications such as portable electronics.

The ion conductivity of a charged specie in a solid solution is proportional to the product of its concentration ($c_i$) and the mobility ($u_i$) of the defect ($\sigma_i \propto c_i \cdot u_i$). Typically, the ionic conductivity has been controlled by doping the oxides with aliovalent cations, which naturally creates charged defects to maintain the electroneutrality, or by trying to stabilize high-conductive crystallographic phases in polymorphic materials. Nonetheless, limitations came across in terms of materials instability, reactivity and defect interactions when trying to implement and operate these ionic conductors in real devices.

A different strategy for the enhancement of the ionic conductivity in a material is based on the use of nanoionic effects, which take place when reducing the dimensions to the nanoscale [1,2]. By using interface-dominated materials, like thin films, both the concentration of charge carriers $c_i$ and the mobility $b_i$ can be modified. Engineering nanostructures can allow tuning these changes, to maximize their impact on the ionic conduction [3,4]. Moreover, interface-dominated materials can also be used to enhance other mass transport phenomena, such as, for instance, the Oxygen Reduction Reaction (ORR) capabilities of a cathode for solid oxide fuel cells [5,6] or the redox reversibility of a material employed in resistive switching [7].

This chapter is aimed to illustrate the most significant nanoionics effects involving interfaces and to explore the possibility of using them for improving the performance of thin film-based devices. An interface is described as an interruption in the symmetry of the crystal structure, which brings strong modifications of the atomic properties observed in the bulk. In section 2, we will theoretically analyse the modification of the defect concentration and/or ion mobility in two types of interfaces, namely, the naturally occurring grain boundaries and the engineered hetero-epitaxial strained interfaces. The third section will focus on the analysis of possible strategies for implementing grain boundary and strain-dominated thin films in real devices. Vertically-oriented columnar nanostructures will be presented for exploiting the special features of grain boundaries in polycrystalline films while strained multilayers and vertically aligned nanocomposites (VANs) will be analysed for integrating strain-enhanced conductivity in epitaxial films. To conclude the chapter, possible applications and prospects of nanoionics concepts will be discussed, focusing on energy and information applications.

## 2. Origin of Nanoionics Effects: Local Defects and Interfaces

The first two sections of this chapter will be devoted to the study of how interfaces alter the defect chemistry of a standard oxide material. In Section 2.1, the analysis of the correlation between composition and structural properties taking place in the proximity of a grain boundary interface is presented. Grain boundaries are known to strongly influence different properties of polycrystalline bulk materials [4] and, more interestingly, to even control the behaviour of interface-dominated nanomaterials such as thin films. The presence of structural defects in the *core* of a grain boundary can modify the surroundings of the interface, finally, altering the behaviour of the bulk. Section 2.2 examines consequences of the accumulation of these defects in the core of a grain boundary. Indeed, the enrichment of charged defects can create a volume, called *space charge region*, in which a concentration profile of the different species is generated for preserving the overall charge neutrality. It should be noticed that the space charge theory detailed here for grain boundaries is general and can be applied to charged interfaces of other nature (for instance between two different materials in a composite [8]). Finally, the last section analyses the effect of the lattice strain on the ionic conductivity. Here, we will focus our attention on the effects of the controlled strain generated by epitaxial growth of thin films on substrates with different lattice parameters [9].

*2.1 Atomistic Picture of a Broken Symmetry: The Grain Boundary*

The large impact of grain boundaries on the behaviour of polycrystalline materials received increasing attention in the last decades, especially due to the interest in structural metals and ceramics [10]. However, a full comprehension of the nature and effects of grain boundaries in functional oxides is still lacking, despite their profound influence in the electrical properties of these materials. It is only lately that computational simulations and high resolution characterization techniques (primarily Atomic Force and Electronic Microscopy) reached the required maturity for significantly contributing to improve the atomistic picture of these grain boundaries.

A grain boundary is an interface composed by two crystals of the same composition joined together. It can be defined by the orientation of the grain boundary plane and the angle of misorientation $\theta$ between the two grain crystallographic principal directions (the rotation angle necessary to bring the grains in the same orientation) [11]. Two grains characterized by mirror symmetry around the boundary plane are named symmetrical, if not asymmetrical. The grain boundary can be described depending on the angle of misorientation $\theta$. This way, the low-$\theta$ angle grain boundary can accommodate the differences in the lattice orientation by an array of dislocations. It is possible to forecast the periodicity of the rearrangement by using the Frank's equation [12], which shows that the dislocation spacing decreases with increasing $\theta$. For high values of misorientation angle the dislocations tend to overlap and several accommodation structures can be originated. One useful way to differenciate the high angle grain boundaries is the so-called *coincidence site lattice* method (CSL) [12,13]. Let us image a grain boundary where two crystals are rotated through the grain boundary plane by an angle $\theta$; in such situation, certain atomic positions from both crystal orientations would coincide, periodically repeating through the grain boundary´s lattice structure. The CSL method uses the reciprocal density of these coincident states, $\Sigma$, to describe any grain boundary. In particular, $\Sigma$ is defined as the ratio between the number of coincident sites and the total number of sites. This way, a grain boundary where two grains with parallel orientations coincide corresponds to Σ=1, while an increasing Σ value represents a less coherent structure.

The high degree of structural disorder found in the different types of grain boundaries can influence on the defect chemistry of the material, leading to a local change in its chemical composition. Many studies have found that oxygen vacancies are created within the firsts atomic planes in several oxides [14–21]. For example, An and co-workers analysed an 8% mol. Yttria stabilized Zirconia (YSZ) Σ13 (510)/[001] symmetric tilt boundary by aberration-corrected TEM and quantified the atomistic composition by Electron Energy Loss Spectroscopy (EELS) [14]. They found that a large oxygen deficiency is detectable in the first 0.5 nm from the grain boundary plane (Figure 1.1). A higher concentration of oxygen vacancies in the proximities of grain boundaries has been found also in other materials such as Ceria [15–17], Strontium Titanate [19] and Barium Titanate [18]. Feng et al. studied the atomic composition of different grain boundary orientations of Ceria by High Angle Annular Dark Field (HAADF) Scanning Transmission Electron Microscopy (STEM) [16]. Interestingly, they discovered that only some orientations present oxygen vacancy accumulation, while others do not show any compositional change. Supported by theoretical calculations, they discovered that the oxygen vacancies were present just in the grain boundaries with the larger degree of structural distortion. They concluded that, by reducing the oxygen content, the most distorted structures could relax while the lattice defects were reduced. Other studies by Liu et al. and by Hoyo et al. also found that the rearrangement of oxygen vacancies is a way through which different crystal structures (e.g. fluorites, perovskites) are able to accommodate the loss of coherency [15,22].

Along with the oxygen vacancies, also cations and dopants can modify their concentration to accommodate lattice defects [23–26]. A recent study by Frechero et al. found that a symmetrical 33° [001] tilt grain boundary in YSZ films not only presents an accumulation of oxygen vacancies, but also of Yttrium [24]. Lee and co-workers also studied the defect distribution around Σ5(310)/[001] grain boundary of YSZ, as well as Ceria, by a hybrid Monte Carlo–molecular dynamics simulations [27]. They discovered that oxygen vacancies are accumulated firstly in the grain boundary due to lower vacancy formation enthalpy, whereas the dopant appears to accumulate as a result of a strong vacancy-dopant interaction. It is to be noted that this interaction can be detrimental for high oxygen diffusion through the grain boundary, because of the large concentration of dopants that can eventually immobilize the charge carriers [26].

These analyses prove that the grain boundary misorientations can locally modify the defect chemistry of the material, providing a way to alter the bulk behaviour. One interesting question that arises is whether the defects generated in this confined region (which extends for approximately 1 nm) are mobile enough to provide a preferential conduction pathway. If so, this could be actively used for modifying bulk properties in the nanoscale. It has been demonstrated in computational studies that dislocations in $SrTiO_3$, although showing an accumulation of oxygen vacancies, do not enhance oxygen diffusion [28,29]. Similar results have been found also for Ceria in [30]. However, recent studies showed that in Strontium doped Lanthanum Manganites the grain boundaries are indeed able to increase oxygen diffusivity by several orders of magnitude [6,31,32]. Further work is therefore necessary to explain the origin of the different behaviours found.

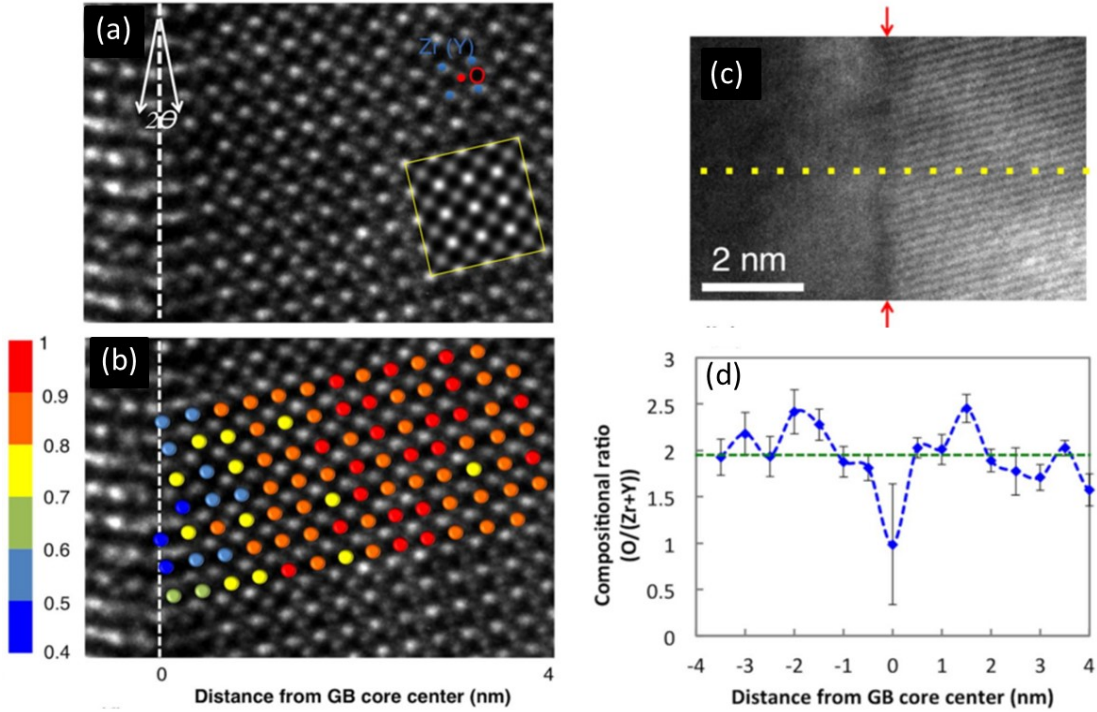

**Figure 2.1.** Aberration-corrected TEM image of a Σ13 (510)/[001] YSZ grain boundary with simulation of crystal in the yellow inset (a) and normalized oxygen column intensity representation (b). STEM image of the grain boundary (c) and variation of O/(Zr+Y) compositional ratio by STEM-EELS. All the figures are reproduced form [14].

*2.2 Space Charge Layer*

*2.2.1 Formation of the Space Charge Layer*

The modification of the defect energy formation in the grain boundary leads to a restricted zone in which charged defects accumulates, i.e. the grain boundary core. Therefore, the electroneutrality is no longer preserved and an electric field is developed in the volume adjacent to the grain boundary plane. In order to keep the charge neutrality, charged species in the surroundings of the grain boundary will tend to modify their concentration, thus compensating the generated charge in the core. The volume of material affected by the electric potential created in the grain boundary and the subsequent redistribution of charged defects is called the *space charge region*. While the grain boundary core takes place in a very restricted zone, *ca.* 1 nm, the space charge region can be extended for even several tens of nanometers. The grain boundary core and the space charge region constitute the "*electrical grain boundary*", which is on the basis of the space charge theory described in the following [4].

Following the simple approach proposed by De Souza for acceptor doped $SrTiO_3$ [33], the space charge region can be presented as a redistribution of mobile charge carriers, e.g. oxygen vacancies for doped $SrTiO_3$, from the bulk to the interface induced by the low defect formation energy at the surface level. In the next, we will develop this analysis to illustrate the space charge layer formation (being aware that a full description of the phenomenon requires a more complex set of hypotheses).

The electrochemical potential $\eta_i$ of a charge carrier subjected to an electrical potential $\Phi$ is defined as [34]:

$$\eta_i = \mu_i + z_i e\Phi = \mu_i^0 + k_b T \cdot ln\frac{c_i}{N_i - c_i} + z_i e\Phi \qquad \text{Eq. 2.1}$$

Where $\mu_i^0$ is the standard chemical potential, $N_i$ and $c_i$ are the number of sites per unit volume and the concentration of defects respectively and $e$, $k_b$ and $T$ are the electron charge, the Boltzmann constant and the temperature. Due to the high defect concentration expected in the core of the surface, which can reach the maximum available sites, the entropy term was defined by using the Fermi-Dirac distribution [35].

Let us consider the formation of a grain boundary by joining together two crystals of the same material but with different orientations, at a temperature high enough to allow the oxygen vacancies to freely move into the lattice. By solving the equilibrium equation $\nabla \eta_v = 0$ between the core and a position $x$ inside the bulk (considering the vacancy migration as a chemical reaction [36]), one can find:

$$\mu_{v,c}^0 - \mu_{v,b}^0 + k_b T \cdot \ln \frac{c_{v,c}}{N_{v,c}-c_{v,c}} - k_b T \cdot \ln \frac{c_{v,b}(x)}{N_{v,b}-c_{v,b}(x)} - 2e\Phi(x) = 0 \qquad \text{Eq. 2.2}$$

The standard chemical potential of the grain boundary core ($\mu_{v,c}^0$) differs from the one of the bulk ($\mu_{v,b}^0$) due to the lower defect formation energy at the surface described in section 2.1. This difference forces a flux of vacancies from the bulk towards the interface, i.e. a redistribution of charged defects in a region close to the interface (space charge layer) and the consequent formation of an electrical potential gradient. Accordingly, vacancies will be transferred to the grain boundary core until its charge ($Q_c$) is fully compensated by a charge of opposite sign in the two sides of the interface ($Q_{sc}$), fixing $Q_c = -2 \cdot Q_{sc}$ (Figure 2.2). In the case of pure accumulation of oxygen vacancies ($z_i$=2) in the boundary core its charge can be calculated as:

$$Q_c = 2ew_0 \cdot (c_{v,c} - c_{v,b}) \qquad \text{Eq. 2.3}$$

With $w_0$ being the core width, usually considered about 1 nm thick. The space charge zone is then characterized by a charge density $\rho(x)$, linked to the core charge by:

$$Q_c = \int_0^\infty \rho(x)\, dx \qquad \text{Eq. 2.4}$$

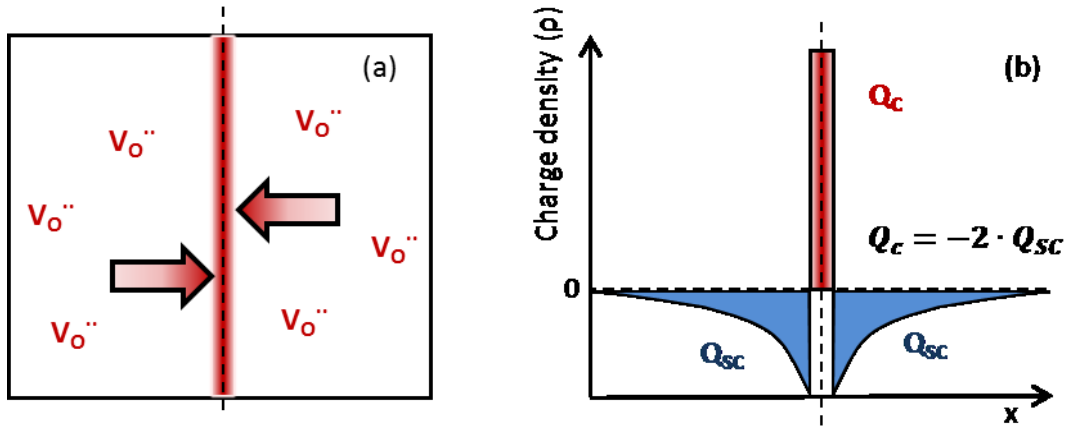

**Figure 2.2.** (a) Redistribution of the oxygen vacancy in a bicrystal due to an easier reducibility of the core and (b) formation of a space charge zone to maintain electro neutrality according to Gouy-Chapman model.

A more complete description of the situation should include the equilibrium with the environment, allowing not just the migration of vacancy inside the material but also any oxygen exchange reaction [15,34]. Additionally, other species besides the oxygen vacancies can accumulate in the grain core or in the space charge region, varying the charge balance and determining more complex and interdependent profiles [23–26].

In order to make space charge theory more general and applicable to any kind of charged interface, the grain boundary behaviour is considered just as a two-dimensional interface with a potential $\phi_0 = \phi(0) - \phi(\infty)$, where $\phi(\infty)$ is the potential in the bulk. Due to the local non-equilibrium, the potential in the space charge layer distributes according to the Poisson equation [4,37–39]:

$$\frac{\partial^2 \Phi}{\partial x^2} = -\frac{1}{\varepsilon_0 \varepsilon_r} \cdot \rho(x) \qquad \text{Eq. 2.5}$$

Being $\varepsilon_0$ and $\varepsilon_r$ respectively the vacuum and material specific dielectric constant. Considering the mobile species as dilute and not interacting, we can obtain their concentration in the space charge region by applying once more equation 2.1. The standard chemical potential will be considered identical in every position, meaning the defect formation energy in the space charge is equal to the bulk one. Moreover, the entropy term can be approximated with a Maxwell-Boltzmann type equation. Thus, the concentration obtained for the *i* species can be expressed as follows:

$$c_i(x) = c_{i,b} \cdot \exp\left(\frac{-z_i e \cdot \Delta\phi(x)}{k_b T}\right) \qquad \text{Eq. 2.6}$$

With $z_i$ being the charge carrier number and $\Delta\phi(x) = \phi(x) - \phi(\infty)$ the potential at the position *x*. The charge density is equal to the sum of the charge of all the individual species ($\rho = \sum_{i=1}^{n} z_i e c_i(x)$), therefore the general expression of the Poisson equation becomes:

$$\frac{\partial^2 \Phi}{\partial x^2} = -\frac{1}{\varepsilon_0 \varepsilon_r} \cdot \sum_{i=1}^{n} z_i e \, c_{i,b} \cdot \exp\left(\frac{-z_i e \cdot \Delta\phi(x)}{k_b T}\right) \qquad \text{Eq. 2.7}$$

Equation 2.6 is general but, unfortunately, there is no analytical solution to it. In order to obtain simple expressions to predict the space charge behaviour, one has to adopt approximations that make the solutions valid under certain restrictions. The different models are based on estimations of the charge density. The principal approximations are the Gouy-Chapman and the Mott-Schottky models [4,37,38,40,41].

The Gouy-Chapman model is a limit solution valid in case of large accumulation of defects in the space charge zone [4,40]. In the model, the charge compensation is considered to be mainly controlled by the dopant cation distribution in the space charge region. Therefore, the charge density can be written as:

$$\rho = -z_c e c_{c,b} \cdot \exp\left(\frac{-z_c e \Delta\phi(x)}{k_b T}\right) \qquad \text{Eq. 2.8}$$

Solving equation 2.5 using semi-infinite boundary conditions and considering large effects [39], one can find:

$$\Delta\phi(x) = \phi_0 + \frac{2k_b T}{z_c e} \ln\left(1 + \frac{x}{2\lambda} \exp\left(\frac{-z_c e \phi_0}{2k_b T}\right)\right) \quad \text{For } x < 2\lambda \quad \lambda = \sqrt{\frac{\varepsilon_0 \varepsilon_r k_b T}{2 z_c^2 e^2 c_c}} \qquad \text{Eq. 2.9}$$

Where $\lambda$ is the Debye length (or screening length), which defines the width of the space charge. This solution is valid just in the case of cations mobile enough to redistribute in the space charge zone and for large values of core charge. The left part of figure 2.3 shows the redistribution of potential and main charged species (cations, oxygen vacancies, electrons and holes) according to this model for Fe-doped $SrTiO_3$ at 600°C and 0.01 bar, considering an interface potential of 0.8 V. It is to be noted that a more general analytical solution exists for the case of a system composed by two mobile ions with opposite sign but same absolute value ($z_i$=-$z_j$). It is called symmetrical Gouy-Chapman and is not restricted to the case of large accumulation effects [4,40].

The Mott-Schottky model considers that the dopant cation is immobile and gives the most predominant contribution to the charge density. This approximation is found to be valid for describing the grain boundary behaviour in many oxides below 1000°C, due to the low diffusion of cations. For example, we note here the studies made by Guo *et al.* on lightly doped Zirconia [41,42] and by Kim *et al.* and by Guo *et al.* on Ceria [40,43,44]. The density charge in the case of constant dopant concentration is:

$$\rho = -z_c e c_{c,b} \qquad \text{Eq. 2.10}$$

The solution of the potential redistribution under this assumption is:

$$\Delta\phi(x) = \phi_0 \left(\frac{x}{\lambda^*} - 1\right)^2 \qquad \text{For } x < \lambda^* \qquad \lambda^* = \lambda \sqrt{\frac{4|z_c|e}{k_b T} \phi_0} \qquad \text{Eq. 2.11}$$

Where $\lambda^*$ is the width of the space charge zone. The space charge redistribution calculated with this model for Fe-doped SrTiO$_3$, under the same conditions of the Gouy-Chapman case, is shown in the right side of Figure 2.3. One can notice that the main difference between the two models is the space charge width. Indeed, the Gouy-Chapman model shows a smaller extension and sharper defects redistribution, due to large screening ability of cation redistribution. Instead, when the dopant concentration is constant the space charge enlarges and becomes dependent on the surface potential. In both cases, a positive potential causes an accumulation of negative charges and a depletion of the positive ones. Under certain conditions, such as the ones chosen intentionally for this example, the concentration of electrons in the space charge can even exceed the holes in an acceptor doped material. The accumulation of electrons in the space charge region has been indeed found for SrTiO$_3$ [45,46] and Ceria oxides [43,44,47]. It is possible to consider the electron enrichment and hole depletion as a downward band bending of both the valence and conduction bands, with a constant Fermi level [33,37]. Another possible effect of space charge in oxides is an increase of electronic transference number (the ratio between electronic and ionic conductivity), due both to the increase of electrons and to a sharper decrease of oxygen vacancies than holes, as found in LaGaO$_3$ [48,49].

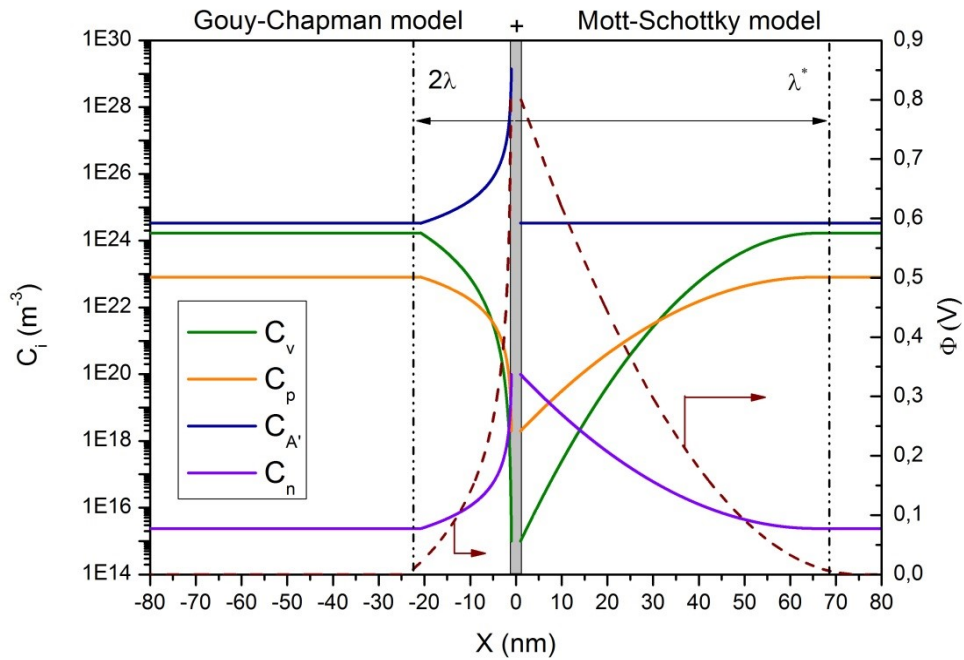

**Figure 2.3.** Gouy-Chapman (left side) and Mott-Schottky (right side) space charge approximation of a Fe-doped SrTiO$_3$ grain boundary calculated for $\Phi_0$=0.8 V, $\varepsilon_r$=160, $C_{A'}$=0.02 atom. % and T=600ºC. The concentration values of oxygen vacancies ($C_v$), holes ($C_p$) and electrons ($C_n$) are derived from the defect chemistry model at 600ºC and 0.1 bar [50,51].

The redistribution of charge carriers causes an observable change in the conductivity in the space charge region (ionic and electronic), since $\sigma_i(x) = |z_i|eu_ic_i(x)$, with $u_i$ as the carrier mobility. The expected ratio between bulk and grain boundary values can be calculated integrating the species concentration along the space charge region, obtaining for the Mott-Schottky model:

$$\frac{\sigma_{i,bulk}}{\sigma_{i,gb}} = \frac{1}{\lambda^*} \cdot \int_0^{\lambda^*} \frac{c_i(x)}{c_b} dx \approx \frac{\exp(ze\phi_0/k_bT)}{2|z|e\phi_0/k_bT} \quad \text{Eq. 2.12}$$

Equation 2.12 can be actually used to obtain the surface potential of a grain boundary if the contributions of grain and grain boundary to the total conductivity are properly distinguished, by, for example, Impedance Spectroscopy [41,43,52,53]. However, the conduction can take place along or across the grain boundary and several different expressions of conductivity variation exists, depending on the space charge approximation adopted [40,43].

When decreasing the grain size to the nanometric scale, the effect of the space charge layer on the overall conductivity becomes more and more relevant, due to the increasing significance of the space charge region[1]. For values of grain size approaching the space charge width, the overall conductivity is expected to change drastically, due to mesoscopic effects [1,3,4,54]. Indeed, for grain size comparable to the Debye length the concentration of mobile defects does not relax to the bulk values, because the space charge of two consecutive grain boundaries overlaps. The change in conductivity is found to vary from the one calculated with a semi-infinite approach by the factor $g$ [4]:

$$g = \left(4 \cdot \frac{\lambda}{d_g}\right) \cdot \left(\frac{c_i(0) - c_i(d_g/2)}{c_i(0)}\right)^{1/2} \quad \text{Eq. 2.13}$$

This mesoscopic behaviour has been found in nanocrystalline SrTiO$_3$ with grain size lower than 100 nm [55]. Due to the increase of electrons concentration, mesoscopic SrTiO$_3$ shows change to n-type behaviour at oxygen partial pressure far higher than the bulk ones. This particular example illustrates the great impact of the nanoionics effects, eventually altering the nature of a material just by controlling the grain size [46].

*2.2.2 Tunability of the Space Charge Layer*

Tuning the properties of oxides by changing the accumulation or depletion of certain charge carriers in the grain boundaries is a fascinating challenge. At this point, it has been made clear that the change in the species concentration along the space charge region depends on the characteristics of the charge core and, consequentially, on the surface potential. It is therefore important to also describe in detail the formation of the electric potential. It has been seen that a positive potential can be generated by the accumulation of positive defects, mainly oxygen vacancies, in the core. This has in fact been found in several different studies for different crystalline structures [14–18]. A more general description of the core charge must however include all other possible local defects, such as cation enrichment (what would provide positive charges in the core) or cation vacancies accumulation (producing a negative contribution to the core charge). The sum of all the contributions will determine the core charge, which for instance in the Mott-Schottky model can be related to the potential height by:

$$|Q_c| = \sqrt{8\varepsilon_0 \varepsilon_r c_{c,b} e |\phi_0|} \qquad \text{Eq. 2.14}$$

---

[1] According the brick layer model, the increase of resistance will depend to the grain size ($d_g$) and the space charge width ($\lambda^*$): $R_{i,gb}/R_{i,b} = \sigma_{i,b}/\sigma_{i,gb} \cdot \lambda^*/d_b$.

Up to now, we have focused on the description of grain boundaries with positively charged cores; in fact, the net charge of the grain boundary core has found to be positive for most perovskites, such as $SrTiO_3$ [33,53,56,57], $LaGaO_3$ [48,49], $BaTiO_3$ [58–60], $BaZrO_3$ [61] and fluorites, such as acceptor doped $ZrO_2$ [14,38,41,42] and $CeO_2$ [40,43,47,62,63]. Few examples of negative core charge can however be found in literature, such as the case of the $TiO_2$. In Titania, a negative surface potential can be developed by the accumulation of cations vacancies in the grain core [64–66]. Unlikely, the sign of the core charge is not easily switchable, because it depends on complex interaction between grain boundary misoriantations and material defect chemistry (see section 2.1).

The width and influence of the space charge is however modifiable, for instance by changing the doping level. Equation 2.11 shows that increasing the cation doping the space charge width decreases, due to a higher capability of charge screening. Furthermore, for large values of doping levels some studies show that other mechanism beyond the space charge depletion are responsible for the observed increase of oxygen conductivity [24,26,67]. The limitation of the space charge model could come from the assumption of dilute and not-interacting species on which is based. Recently, Mebane and co-workers developed a model based on the Cahn–Hilliard theory for inhomogeneous systems [68]. They found that in Gadolinia-doped $CeO_2$ for values of dopant higher than 1% mol., the model forecast a segregation of Gadolinium along with a complex behaviour of oxygen vacancies that experience both depletion and enrichment through the space charge zone. This model was found to agree with atomistic simulations that emphasise the role of vacancy-dopant interaction in the grain boundary [27]. While for low dopants values the classical space charge theory can effectively predict the grain boundary behaviour, further work is necessary to fully understand the properties of heavily doped materials.

From the considerations drafted in this section, one can expect predominant space charge behaviour from materials characterized by large dielectric constant and small doping levels. Increasing the doping level will reduce the space charge extent and give rise to interactions dopant-defect. Therefore, for obtaining large nanoionics effects, thin films with a low concentration of point defects and nanometric grain size should be explored.

2.3 *The strain effect on ion mobility*

In oxide thin films with a large amount of point defects, such as Gadolina-doped Ceria (GDC) and YSZ, phenomena associated to the change of defect formation enthalpy and of defect concentration are not expected to play a significant role in the increase of oxygen conductivity. This is mainly because the charge carrier concentration is not limiting the diffusion anymore but their mobility. For instance, in Zirconium oxides the oxygen vacancies can be increased arbitrarily by doping with a trivalent cation such as Yttrium, but the maximum in conductivity is found for doping values around 8% mol. of $Y_2O_3$. Higher concentrations reduce the mobility of the oxygen vacancies due to strong interactions with the Yttrium dopant. In addition to that, it is important to highlight that small space charge effects are expected in such heavily doped oxides, since the space charge width is inversely proportional to the doping level, see Equation 2.9 [67]. Nevertheless, interfaces can still offer a way to improve the oxygen conductivity in these materials, taking advantage of the correlation between lattice strain and oxygen ion mobility [9,69–75].

The literature on ionic conduction in oxide thin films is vast and shows a wide scattering of values for similar materials in different works [70]. Improvement as well as diminishment of the ionic conduction is found for even the same material. In particular, the literature on YSZ in references [74,76–78] present a great variability and no clear correlation of the results with the fabrication conditions. Although some artifacts in the measurements can arise when nanometric thin films are measured [79], it has been proposed that lattice strain can deeply influences ionic diffusion [70,73]. Barriocanal and co-workers reported a *colossal* enhancement of ion conductivity in YSZ:STO epitaxial heterostructures grown on STO substrate [75], showing an improvement by 7 orders of magnitude in the oxide ion conduction. In their work, they claimed that the origin of this improvement was associated to the highly strained interface of YSZ films (the lattice mismatch lead to a 7% of expansion), creating planes with high mobility and a large number of carriers. The origin of this *colossal* ionic conductivity is however still debatable, and successive studies have pointed out that the obtained conductivity could have an electronic origin, due to the p-type electronic conductivity found on STO surfaces [80,81]. Still, although the origin of the YSZ/STO high conductivity is still unclear [82–84], this work inspired many other studies for trying to tailor the strain effect in oxide ion conductors. For example, Sillassen *et al.* showed that conductivity of YSZ deposited on top of single crystal MgO increases when decreasing the film thickness, which might be related to an interfacial zone full of lattice defects [74]. Similar results were found for YSZ deposited on Sapphire, where the tensile plane strain was able to reduce the energy of activation of the conduction process [76]. Korte and co-workers, through a series of different studies, analysed the effect of strained YSZ interfaces growing oriented multilayers of YSZ and one insulating oxide that put the YSZ layer under strain [85–87]. They demonstrated that tensile strain is able to enhance ion mobility in YSZ, although they found lower conductivity enhancements with respect to the ones found by Barriocanal *et al.* [75]. Effect of compressive stress on the conductivity was also recently shown on GDC free-standing membranes [88]. In this work, higher activation energies were however found on the released membranes when compared to the substrate-supported strained films.

To better understand the phenomena, several theoretical studies have been performed to investigate the effects of lattice strain in the conduction mechanism of Yttria stabilized Zirconia. Araki and co-workers studied the influence of uniaxial strain on the oxygen ion conductivity using Molecular Dynamics simulations [89]. In the direction of the applied strain they reported an increase of diffusivity in the case of tensile stress and a decrease when applying compressive stress. They concluded that the elastic strain contributed to enhance or decrease the oxygen diffusion in the direction of the applied force. Kushima *et al.* studied the oxygen vacancy migration paths and barriers in YSZ as a function of lattice biaxial strain using density functional theory (DFT) and nudged elastic band (NEB) method [90]. They found that tensile strain can improve ion diffusion up to 2 orders of magnitude at 800 K by increasing the migration space of oxygen in the lattice and by reducing the bonding strenght between oxygen and cations. Other molecular dynamics studies demostrated that tensile strain can decrease the energy of activation of the diffusion process, while compressive strain increases it [91,92]. Particularly, Tarancón *et al.* found that under tensile strain the activation energy decreases till values of 3% and then increases dramatically [92]. They discovered that this is due to a structural relaxation of the lattice that leads to an increase of relaxation enthalpy and the formation of a new equilibrium position for the oxygen (Fig. 2.4). Although there is a certain divergence in the quantification of the oxygen conductivity variation under strain, all the studies show agreement with the experimental findings that tensile strain enhance oxygen ion mobility while compressive strain hinders it.

In addition to the atomistic simulation, an analytic qualitative approach to forecast the behaviour of strained oxide interfaces in thin films has been developed by Korte and co-workers [9,85,86,93,94]. The model is based on the relation between isostatic pressure (derived from the strained interface) and defect migration enthalpy [9,95]. The relation obtained is:

$$ln\frac{\sigma_{V_{\ddot{O}},strain}}{\sigma_{V_{\ddot{O}},bulk}} \approx \frac{1}{3}\frac{\Delta V_{V_{\ddot{O}}}}{RT}\cdot\frac{E}{1-\nu}f_{1,2}\qquad\text{Eq. 2.15}$$

Where E is the Young module, $\nu$ is the Poisson coefficient and $f_{1,2}$ is the lattice mismatch[2] between phase 1 and 2. $\Delta V_{V_{\ddot{O}}}$ is the migration volume of the oxygen vacancy, which is positive and has been estimated around 2.08 cm$^3$/mol [96]. The relation predicts an enhancement of conductivity at 500°C of almost 2 orders of magnitude for a strain of 2%. Still, this relation just holds in the case of elastic strain. In the event of high lattice mismatch, the structure is known to accommodate the defects by dislocations and structural defects, therefore equation 2.15 is not valid anymore [69].

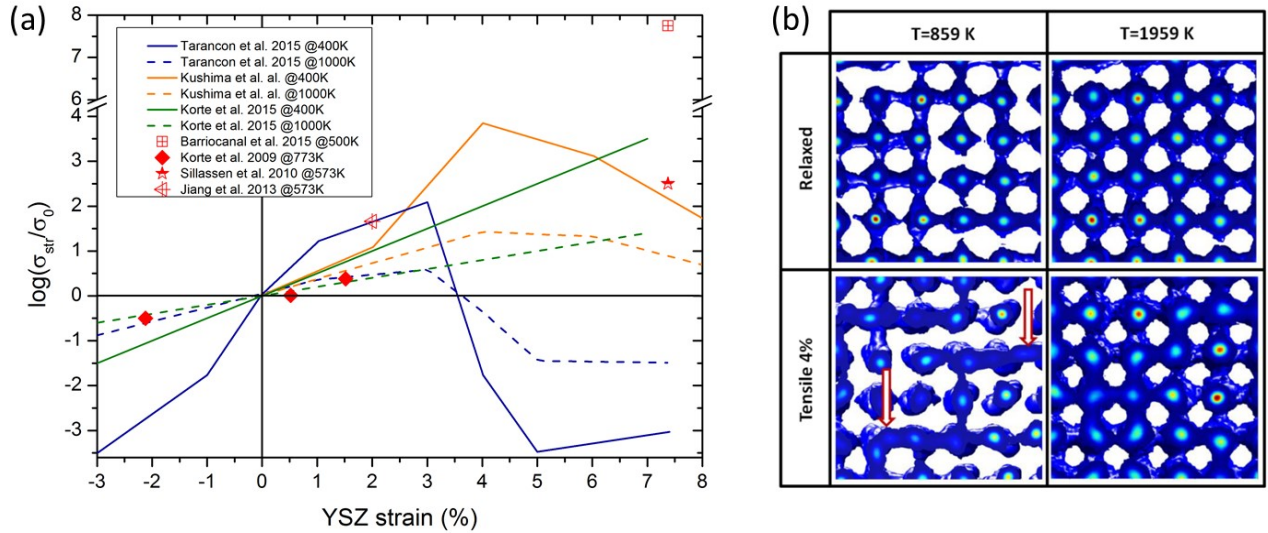

**Figure 2.4**: Variation of the oxygen conductivity in YSZ with elastic according to simulation results of Tarancón *et al.* [92], Kushima *et al.* [90] and to the model developed by Korte et al. [9,85,86,93,94] at 400K and 1000K; the experimental values founded by Barriocanal *et al.* [75], Koerte *et al.* [9], Sillassen *et al* [74] and Jiang *et al.*[76]. For strain higher than 3% Tarancón *et al.* observed a relaxation of the structures, with oxygen atoms stable in interstitial position. The atomic density map of the oxygen sublattice (b) is reproduced from [92].

It is important to notice that grain boundaries and, in particular, dislocations, are extremely connected with high strain fields. When the elastic strain is too high to be accommodate by lattice mismatch, a lattice defect can be formed [92]. However, although the overall strain is reduced, high local opposed strains are generated in the proximities of the defect. Recent studies by Yildiz and co-workers have shown how elastic strain around a dislocation influences the energy formation of oxygen vacancies [29]. Overall, a grain boundary can ultimately be seen as an array of dislocations. Therefore, strong correlations are expected between the two phenomena.

Strained interphases appear as the most promising solution for increasing oxygen conduction in pure ionic conductors. Moreover, lattice strain is thought to be responsible for other phenomena in oxides thin films, such as oxygen vacancy formation and improvement of oxygen exchange reactions [71]. Therefore, lattice strain can improve the ionic properties of many micro solid-state electrochemical devices, such as µSOFC [72].

---

[2] The lattice mismatch $f_{1,2}$ is distributed in the strain of phase 1 and 2, accordingly to: $f_{1,2} = \varepsilon_1 - \varepsilon_2$. If the two phases have similar mechanical properties, the approximation $\varepsilon_1 = -\varepsilon_2$ is valid and the lattice mismatch become: $f_{1,2} = 2*\varepsilon_1 = 2*\varepsilon_2$

## 3. Strategies for the implementation of nanoionics in functional oxide thin films

Following the description given in the previous sections, two main strategies are identified for profiting from nanoionic effects in real devices, namely, the use of grain boundary-dominated materials and the integration of strain-tuned epitaxial films. In this section, we will go deeper into these two approaches and will provide key literature references where such nanoionic effects are shown in functional oxide films.

*3.1 Grain boundary-dominated materials*

Typically, the active use of grain boundaries for modifying the functionality of materials has been found difficult to implement, despite polycrystalline materials being vastly deployed in bulky systems and also in many thin film-based devices. The random orientation of the grains and their relatively big size (compared to the grain boundary width) have contributed to classically see the grain boundaries rather as a *problem* than a *solution* [3,24]. In the particular case of ionic diffusion, this view is mainly related to the lower diffusion observed for the mobile ions *across* a grain boundary core [97]. However, the great advances in thin film deposition and characterization technologies attained in the last decades opened new possibilities to actively use grain boundaries for improving the performance of certain materials. Thin films with reduced dimensions and customized microstructures (grain size, density, crystallinity…) can be fabricated with high reproducibility, thus making grain boundaries accessible not only for their study and better understanding [14,17,21,23,98] but also for actively profiting from their modified properties [5,32,99]. For example, thin films of functional oxides can be deposited by pulsed laser deposition (PLD) in the form of fully packed vertically aligned columnar grains [6]. This configuration opens a new avenue for novel devices and applications by allowing the use of enhanced properties of grain boundaries *through* the films in vertical configurations.

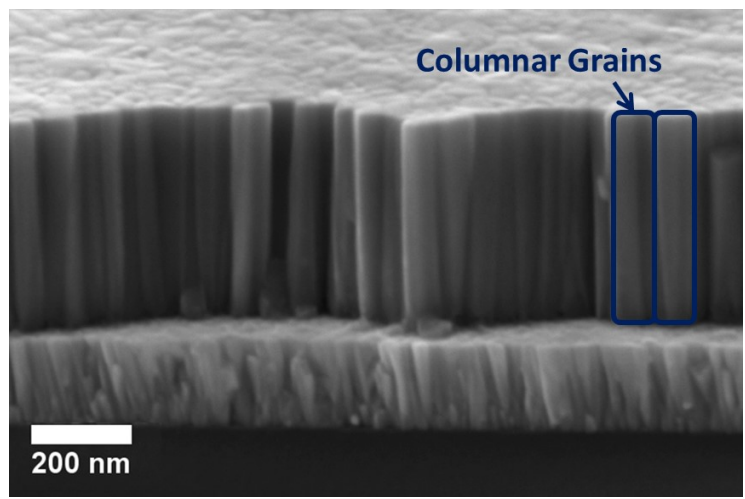

**Figure 3.1.** Columnar-oriented thin film of $Sr_2Fe_{1.5}Mo_{0.5}O_{6-\delta}$ grown by Pulsed Laser Deposition at Catalonia Institute for Energy Research, Barcelona, Spain.

We have seen in previous sections that grain boundaries effectively affect the charge carrier concentration in the grain boundary core and its surroundings. Based on the depletion or enhancement of different charge carriers in either the grain boundary core or the space-charge region, different possibilities are foreseen for grain boundary engineering.

First, tailoring grain boundaries could help to *maximize ionic conductivity*. No clear evidences of enhanced ionic conduction along the grain boundaries has been found in classical ionic conductors such as YSZ [100] or ceria [101]. Instead, strong immobilization of charge carriers might happen and conduction is lowered not only across the grain boundary but also along it. Nanostructuring the oxide thin films can however still benefit the overall ionic conduction, precisely by minimizing the negative effect of grain boundaries. In polycrystalline films, columnar grains are in this case useful for ion conduction through the film, since the number of crossed interfaces will be minimized or even neglected. In this sense, recent results on polycrystalline YSZ films with vertically aligned nanograins showed an associated resistance only attributable to the bulk part [102]. Since these films were fabricated as membranes, access to both sides of the film is possible, allowing multiple real implementations for this enhancement, e.g. micro-SOFCs. This strategy makes easy-to-implement polycrystalline films almost equivalent to epitaxial films in terms of performance. Yet, it is remarkable to mention that recent results in some particular materials have also shown an enhancement of ion mobility in the grain boundaries. It is the case of strontium-doped lanthanum manganite (LSM) mixed ionic-electronic conductors. Independent works from Fleig's and Tarancón's groups recently discovered fast oxide-ionic conductivity along grain boundaries in LSM, reporting up to 5-6 orders of magnitude of improvement in oxide-ion diffusivity compared to the bulk, which might be associated to strain-induced defects and corresponding changes in the oxidation states of LSM constituent cations [6,31]. In this sense, dominating ionic conductivity at the grain boundary level was reported for PLD-growth thin films of $La_{0.8}Sr_{0.2}Mn_{1-x}Co_xO_{3\pm\delta}$ [103], $LaCo_{0.6}Ni_{0.4}O_3$ [104] and consistently from Develos-Bagarinao *et al*. for YSZ, CGO, $La_{0.6}Sr_{0.4}Co_{0.2}Fe_{0.8}O_{3-\delta}$ and $La_{0.6}Sr_{0.4}CoO_{3-\delta}$ [105,106].

Grain boundaries could be also used for *enhancing electrochemical activity*. Assuming the formation of a positively charged grain boundary core (observed in most of the metal oxide thin films analysed up to now, see section 2), a reduction of vacancy concentration is expected in the space charge region in order to keep the charge neutrality. The higher concentration of oxide ions in this zone can therefore influence on the oxygen reduction kinetics, as well as surface electronic structure (charge transfer). The distinct oxygen diffusion and surface exchange kinetics in the grain boundaries was analytically described by Preis *et al*. [107,108]. Experimentally, Shim *et al*. found evidences of an enhanced oxygen reduction reaction rates in YSZ surfaces with smaller grains, thus suggesting the positive effect of grain boundaries in enhancing surface reactions in YSZ. Lee *et al*. also studied the oxygen surface exchange at grain boundaries in GDC [5]. They concluded that higher grain boundary densities in GDC results in a higher exchange current density, evidencing the effect of grain boundaries on surface kinetics. Importantly, they also showed an interesting analysis of the charge distribution near the grain boundaries measured by Kelvin probe microscopy, finding a positively charged core for GDC. Doria *et al*. also showed higher oxygen vacancy concentrations and slower diffusion in the grain boundaries of SDC thin films [109]. Finally, evidences of enhanced electrochemical activity has been also found in the grain boundaries of strontium-doped lanthanum manganite (LSM) directly related to the previously mentioned enhancement of the ionic conductivity along grain boundaries [32]. According to the oxygen exchange model proposed by Chiabrera *et al*. for interface-dominated LSM [31], the origin of this enhancement is found in the presence of a high oxygen vacancy concentration at the grain boundary level in LSM thin films opposite to the limited amount of available vacancies typically occurring in LSM bulk.

Finally, it is foreseen that engineering grain boundaries could help in controlling redox capabilities at the interface level. The use of nanoionic effects in metal oxide thin films have been indeed already proposed for the fabrication of new resistive switching functional films based on interfacial oxidation/reduction of MIECs [110]. In particular, the modified electronic concentration and band structure in the grain boundaries could potentially help to separate in the nanoscale ionic and electronic channels at high density. However, as highlighted by Messerschmitt *et al*., the specific role of thin film microstructural properties is still not fully understood, and further work is needed for understanding the effects of grain boundary density, and also strain, in the resistive switching mechanisms [111].

*3.2 Strained epitaxial films*

The effectiveness of strain tuning in altering electrochemical properties (mainly ion mobility) in oxide thin films has been deeply analysed by many computational studies and also recently confirmed by analogous experimental works (see section 2 in this chapter). In general, epitaxial films can be grown on single crystal materials and strain is generated in the films due to the lattice mismatch between film and substrate. The classical approach for studying strain effects in the thin films properties has been using the so-called substrate-strained films, i.e. elastic strained lattices directly formed on a single crystal material. However, the ability of effectively strain films by substrate lattice mismatch is limited. A critical thickness in which elastic strain is maintained is usually between a few nm and ~50 nm, depending on the mismatch. With increasing thicknesses, the film crystal structure tends to relax towards the stress-free state typically by the formation of dislocations. Because of that, the applicability of these substrate-strained structures as functional films in real electrochemical devices has been found typically limited.

In order to expand the strain effects to more applicable dimensions, two complementary strategies have been pursued, namely, i) the deposition of epitaxially-grown strained multilayer materials and ii) the fabrication of epitaxial vertically aligned nanocomposite films (VANs). Both approaches target to increase the thickness of the strained film to more practical values, but they differ in the effective direction of the generated strain. While by creating multilayers planar strained lattices are generated (strained pathways formed in the xy plane parallel to the substrate), through the deposition of VAN films a strained lattice is created perpendicular to the substrate; i.e. the modified film properties derived from the strained lattice can be accessed on a cross-plane configuration. The complementarity of these two approaches is crucial, since it potentially allows the implementation of strain-engineered films in devices with different designs. In the following, the effectiveness of both strategies for film strain tuning is described and key literature references are provided.

<u>Multilayers.</u> Many efforts have been derived lately to the study of strained multilayers as a way to increase mobility of charged species. In general, the use of strain to alter ion mobility is based on the idea that the local weakening of bonds in the form of elastic stretching should enable more facile oxygen migration [71]. In the previous section, it was already introduced the astonishing results obtained by Barriocanal *et al.* on the oxide ionic conductivity in YSZ/STO heterostructures [75]. Although subsequent works have put in doubt whether the origin of the 7 orders of magnitude higher (*colossal*) conductivity measured there has either an ionic (through the YSZ) or electronic (through the STO) origin [80,81], it is also clear not only from these but also from other similar works that tensile elastic strain can positively influence the ionic conductivity in heterostructures [69,85,86,93,94,112]. It is however worth to mention that other works from Pergolesi *et al.* [113] or Shen *et al.* [114] showed no significant effects in ion mobility on YSZ and ceria-based multilayers. These works highlighted the difficulties found in obtaining a controllable elastic strain in functional heterostructures, and the probable formation of incoherent interface structures that may relax the lattice, reducing the positive effect of elastic strain. Strain can also affect the concentration and mobility of oxygen vacancies on the surface. Ultimately, this could influence on the oxygen reduction kinetics, as well as surface electronic structure (charge transfer). Kubicek *et al.* studied the effect of elastic strain in altering the oxygen surface exchange and diffusion in epitaxially grown strontium-doped lanthanum cobaltite (LSC), confirming that tensile lattice strain can effectively enhance the oxygen surface reactivity [115]. The influence of strain on oxygen surface exchange and transport was also found in other epitaxially grown materials [116–119]. Finally, another possible application of strained multilayers is the fabrication of resistive switching devices.

Recently, Schweiger *et al*. have developed a "microdot" $Gd_{0.1}Ce_{0.9}O_{2-\delta}/Er_2O_3$ multilayer device by microfabrication process [7,120]. In their work, they showed the possibility of taking advantage of a strained interface for tuning the oxygen transport and, at the same time, of creating a small device able to reach high levels local electric field, necessary for obtaining the non-linear current-voltage behaviour at the base of the resistive switching phenomena.

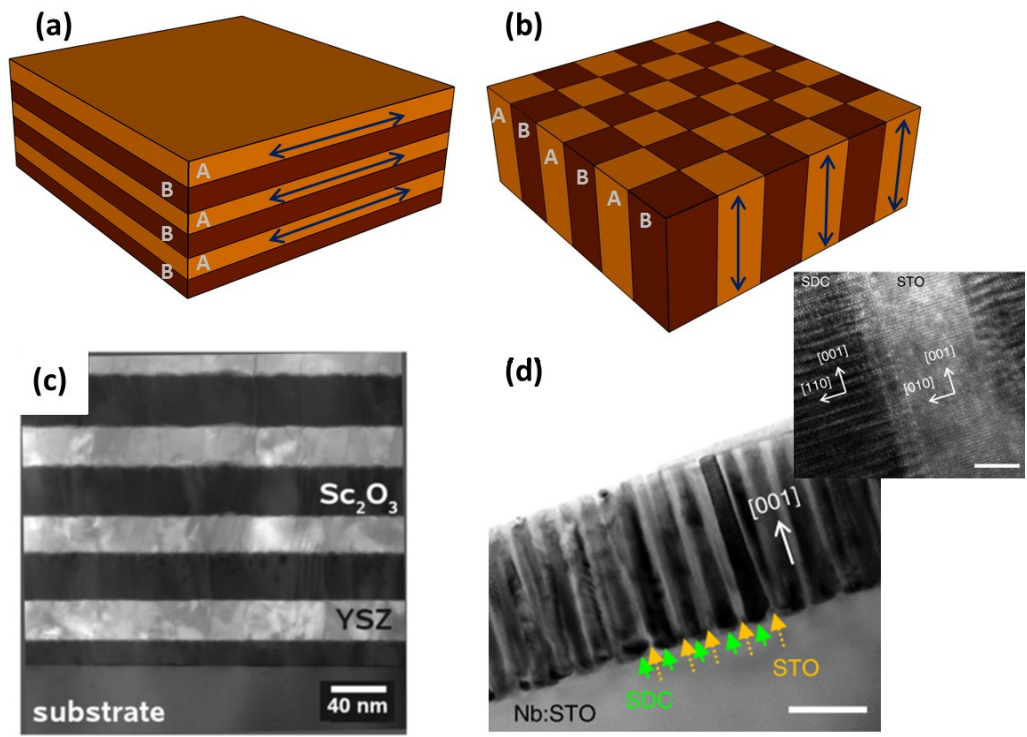

**Figure 3.2**. Sketch of (a) a multilayer epitaxial structure and (b) a VAN structure. (c) TEM images of multilayer YSZ/$Sc_2O_3$ multilayer produced by Korte *et al* [86] and (d) of a Sm-doped $CeO_2$/STO VAN produced by Yang *et al*. [121].

<u>VAN thin films.</u> Recent work by MacManus-Driscoll and co-workers has proven the possibility of epitaxially grow vertically aligned composites in many different systems, and for many different applications, with apparent no limitation in thickness. The advantages of VAN thin films come not only from the possibility of accessing vertically strained functional films for cross-plane configuration devices, but also from the fact of having to epitaxially grown materials in a perfectly oriented thin film. New possibilities are therefore foreseen for the fabrication of multifunctional devices using the distinct properties of each strained material on the nanocomposite. The possible fields of applications of VAN strained films are therefore numerous, including magnetoresistance [122–124], ionic/electronic conductors [121,123,125,126] or resistive switching [127]. VAN films have been also used for fabricating singular nanostructures, like nanoporous epitaxial films [128]. On the other hand, it is important to highlight that nowadays the fabrication of VAN thin films is still restricted to a limited number of materials' combinations. As described by MacManus *et al*. in [129], the materials selection has to follow certain guidelines. In brief, these can be summarized as: the chosen phases should be chemically and thermodynamically stable, they should allow epitaxial growth on a given substrate under similar conditions, and the materials should have different elastic constants in order to create strained interfaces.

One relevant issue for the application of multilayer strained structures is the choice of a proper substrate to grow epitaxial thin films. If not wisely chosen, the substrate can hinder the implementation of the strained epitaxial films in the final device. Moreover, this technical limitation is complemented with a techno-economical weakness based on the use of substrates not considered scalable, i.e. not massively employed in any current industry. One strategy to overcome this obstacle is the utilization of an epitaxial buffer layer to reduce the lattice mismatch between the desired thin film and a custom substrate. For example, Sanna *et al*. succeeded in growing epitaxial Samaria doped Ceria (SDC)/YSZ thin films on MgO, by depositing a buffer STO functional layer between the SDC/YSZ and the MgO [112]. Currently, different research actions are also taken in order to achieve industrial fabrication of relevant epitaxial layers, e.g. STO, on top of scalable substrates used in the microelectronics industry, i.e. Silicon [130].

## 4. Prospects for applications of nanoionics in metal oxide thin film-based devices for energy and information applications

The integration into real devices of functional materials in which nanoionic effects can play an active role is just being explored. Two main fields of application are identified where metal oxide thin films are key components, namely, new energy sources for portable applications and information storage. In particular, there is an increasing interest in the last years on the development of micro solid oxide fuel cells (µSOFC) and all-solid-state microbatteries as energy conversion and storage devices, respectively, for covering the low-to-intermediate power source regimes. At the same time, the fabrication of resistive switching devices based on metal oxide films is receiving increasing attention as new alternative information memories and logics. Figure 4.1 exemplifies the novel concepts porposed for the miniaturization of these three distinct devices, in which the fine control of nanoionics can largely benefit the performance.

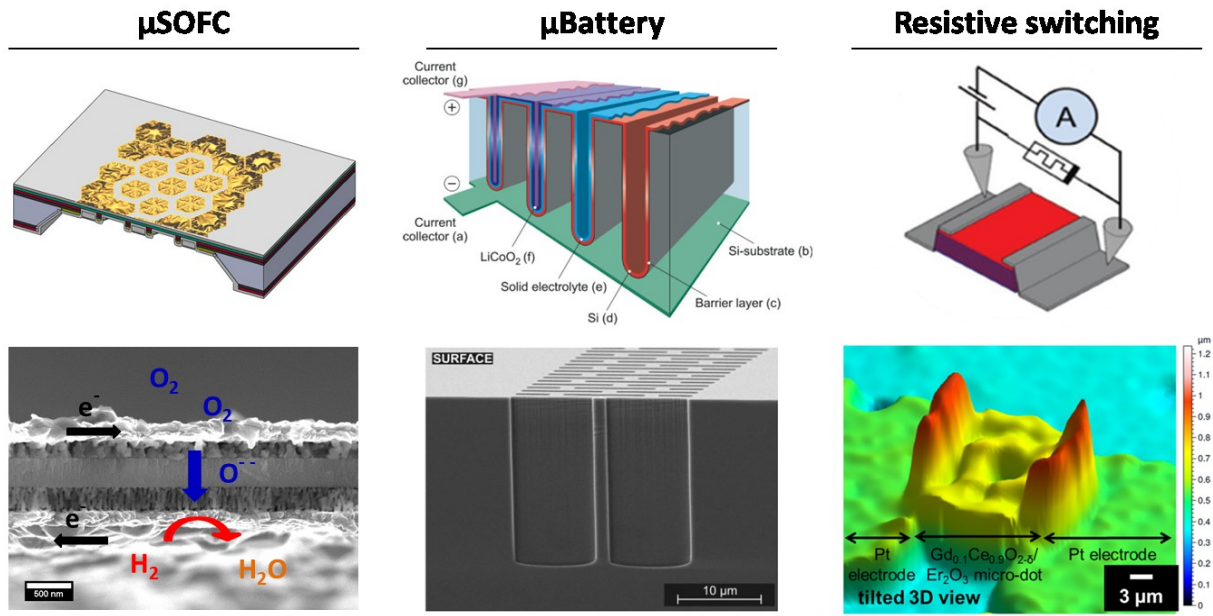

**Figure 4.1**: Literature examples of metal oxide thin film based devices for (a) µSOFC [131], (b) µBattery [132], (c) Resistive Switching [7,120].

In metal oxide-based energy conversion and storage devices, two main issues are identified that strongly limit the final performance, namely, the high internal ionic resistance through the solid electrolyte (oxide-ionic in µSOFCs while Lithium-cationic in microbatteries) and the oxygen surface reactivity and Li$^+$ incorporation in µSOFC and microbattery electrodes, respectively. In the previous sections, we have already seen how nanoionics can actively influence these properties by controlling the defect concentration and mobility of the charge carriers. In particular, grain-boundary and strain engineering can increase the conductivity in thin films for convenient architectures. At the same time, the surface kinetics in electrode films can be equally stimulated by engineering surface defects distributions and electronic conduction at the interface. All in all, the aim of any nanoionics engineering of these multilayer devices is always reducing the total resistance at a certain operation temperature to increase the generated power.

In silicon-integrated μSOFC, free-standing electrolytic membranes are the core element of the device [133]. They play a key role on separating oxidizing and reducing atmospheres, but equally important is the oxygen conduction through it. Using thin films as electrolytes allows a significant reduction of the internal resistance or, alternatively, a strong decrease of the operation temperature. Temperatures below 350ºC are proved for classical materials such as YSZ [134]. Garbayo *et al.* reported a reduction of one order of magnitude of the membrane-through resistance (compared to in-plane resistance) for thin membranes of YSZ made of columnar-oriented nanostructures [102]. In such a vertical configuration, VAN thin films could be also useful if a significant increment in ion mobility was achieved with the vertical strain. Strained thin films and, in particular, VAN configurations could be also used for enhancing the electrochemical activity in μSOFC electrodes, by altering the oxygen surface exchange and diffusion. However, the modified defect chemistry in grain boundaries of dense polycrystalline mixed ionic-electronic conductors may be here the most promising approach for fabrication dense, stable electrodes with high electrochemical activity in large volumes of the material. Recent progress, mentioned in the previous section, in materials for SOFC cathodes (LSM, LSCF, LSMC or $LaNiO_3$) indicate the high potential of this approach.

A similar analysis as for the case of oxide ion conduction in metal oxides can be done for $Li^+$ conducting materials for all-solid state Li microbatteries [135]. In general, the main challenge for the implementation of Li-based solid electrolytes in batteries is the lower conductivity observed, especially compared to the state-of-the-art liquid ones. Moreover, it has been found that Li conductivity across the grain boundaries is typically lower than in the bulk (up to 4 orders of magnitude in the case of the $Li_{3x}La_{0.67-x}TiO_3$ perovskite) and represents the practical limitation for the deployment of this safer technology [136–138]. Here, columnar-oriented thin films could help to diminish the total resistance, by both reducing the thickness and minimizing the negative effect of grain boundaries on the conductivity. One can envision as well that the modification of defect concentration at the interfaces, as well as the effect of elastic strain, could offer new possibilities not explored yet for altering the $Li^+$ resistivity in solid state microbatteries. As a matter of fact, Haruyama et al. has shown that surface modification by the interposition of $LiNbO_3$ in the $LiCoO_2/β-Li_3PS_4$ cathode/electrolyte interface can actively reduce the $Li^+$ depletion at the space charge region, providing smooth Li transport paths [139]. At the same time, although the low dimensions of the active components of the microbattery foresee low capacity values, nanoionics can potentially play a key role on improving the $Li^+$ incorporation in electrodes for fast exchange rates, crucial for enhancing the specific power of thin film solid state battery devices [140]. This feature becomes extremely relevant for applications in which the battery is coupled to a harvester, e.g. in the field of the Internet of Things.

The modified defect chemistry in the interfaces and potential distinct conduction mechanisms make possible to envisage new possibilities for resistive switching devices too [127]. Local accumulation of defects, as well as the appearance of distinct conduction pathways through the oxide films can be actively used for creating multiresistive states, depending on the mechanism. We exemplify this here by imaging a resistive switching device based on a polycrystalline columnar-oriented metal oxide film, in which positively charged grain boundaries form a space charge region through which electronic mobility is greatly enhanced. In such a system, four different resistive states could be possible, i.e. two additional intermediate states could be formed in which either the electron conduction through the grain boundaries or the ion conduction through the bulk is blocked. Moreover, devices based on nanoionics could offer the possibility of implementing in the same structure multiple functionalities, e.g. serving as switch, capacitor and/or artificial synapses [8,141,142].

Last but not least, in this chapter we have described the positive effects and potentialities of nanoionics on the improvement of electrochemistry in metal oxide thin films. It is however fare to mention as well the possible negative effects in downscaling devices to the nanometric range. For example, extrapolating grain boundary oxygen diffusion values calculated for LSM by Saranya *et al*. [6], it is estimated that oxygen may have diffusion lenghts of the order of ~100 nm at temperatures as low as 150ºC in the expected lifetime of a device (10 years), i.e. larger lengths than the film thickness. This may have strong implications in applications where controlled diffusion is needed (e.g. magnetoresistance). Accelerated ageing effects were for example found by Šimkevičius et al. on $La_{0.83}Sr_{0.17}MnO_3$ films subjected to Ar atmospheres at moderated temperatures, which was associated to oxygen depletion in the grain boundaries [143].

To conclude, we have reviewed that local defects changes the electrochemical properties of metal oxide thin films, giving rise to nanoionic effects. Based on that, we have shown the different strategies that can be implemented for actively control the interaction of interfaces and ionic conductivity. This opens new routes for the fabrication of optimized metal-oxide based devices through the engineering of nanoionics, for a new generation of energy and information technologies.